\documentclass[twocolumn,floatfix,notitlepage,groupaddress,showpacs,nofootinbib]{revtex4-1}
\usepackage{times,color,amsfonts,amssymb,amsbsy,amsthm,amsmath,graphics,graphicx,hyperref,bm,bbm,makecell,mathtools,dsfont,upgreek,fancyhdr,multirow,newtxmath,subfigure}
\usepackage{tcolorbox}

\DeclareFontFamily{OT1}{pzc}{}
\DeclareFontShape{OT1}{pzc}{m}{it}%
{<-> s * [1.25] pzcmi7t}{}
\DeclareMathAlphabet{\mathpzc}{OT1}{pzc}%
{m}{it}
\usepackage{tcolorbox} 
\hypersetup{colorlinks,linkcolor={black},citecolor={black},urlcolor={black}}
\let\oldsqrt\sqrt
\def\sqrt{\mathpalette\DHLhksqrt}
\def\DHLhksqrt#1#2{%
\setbox0=\hbox{$#1\oldsqrt{#2\,}$}\dimen0=\ht0
\advance\dimen0-0.2\ht0
\setbox2=\hbox{\vrule height\ht0 depth -\dimen0}%
{\box0\lower0.4pt\box2}}
\definecolor{azure}{rgb}{0.0, 0.5, 1.0}

\usepackage{academicons}
\definecolor{orcidlogocol}{HTML}{A6CE39}
\usepackage{lineno}

\begin{document}

\title{Quantum communications in a moderate-to-strong turbulent space}
\author{Masoud Ghalaii}
\author{Stefano Pirandola}
\affiliation{Department of Computer Science, University of York, York YO10 5GH, United Kingdom}

\begin{abstract} 
{\bf Abstract.}  
Since the invention of the laser in the 60s, one of the most fundamental communication channels has been the free-space optical channel. For this type of channel, a number of effects generally need to be considered, including diffraction, refraction, atmospheric extinction, pointing errors and, most importantly, turbulence. Because of all these adverse features, the free-space optical (FSO) channel is more difficult to study than a stable fiber-based link. For the same reasons, only recently it has been possible to establish the ultimate performances achievable in quantum communications via free-space channels, together with practical rates for continuous variable (CV) quantum key distribution (QKD). Differently from previous literature, mainly focused on the regime of weak turbulence, this work considers the FSO channel in the more challenging regime of moderate-to-strong turbulence, where effects of beam widening and breaking are more important than beam wandering. This regime may occur in long-distance free-space links on the ground, in uplink to high-altitude platform systems (HAPS) and, more interestingly, in downlink from near-horizon satellites. In such a regime we rigorously investigate ultimate limits for quantum communications and show that composable keys can be extracted using CV-QKD. 
\end{abstract}

\flushbottom
\maketitle
\thispagestyle{empty}

\bigskip
{\bf Introduction} 

Yearslong chain of excellent work have stitched quantum communications and quantum cryptography into the science of quantum information technologies. In particular, QKD \cite{Pirandola:AQCrypt} has been developing rapidly, with the end goal of making distant individuals able to share a key, which must be inscrutable for an eavesdropper to learn about, and which, therefore, can be used for secure classical communications. Since 1980s that saw the d\'ebut of QKD \cite{BB84}, optical fibers have been the main platform to perform and/or experiment most QKD protocols. However, the reach of fiber-based quantum communications is limited to only a few hundreds of kilometers \cite{Chen:PRL509km,Zhang:Optica2018,Lucamarini:TF2018,Pittaluga:2020} (because of the exponential decay of the transmissivity). Whereas, man seems to stand on the verge of  building a quantum internet \cite{Kimble:QuInternet,Pirandola:QInternet} to make global quantum communications viable. 

As a possible solution, one may think of a harmonized use of quantum repeater stations (placed on ground and connected via optical fibers) and free-space communication links. The latter includes ground-to-ground free-space channels, HAPSs, downlink/uplink communications with satellites, and inter-satellite links.  
{To make secure free-space and satellite QKD globally available, certain technological challenges must be addressed. 
There has been increasing attempts put by the community in this direction; many models have been proposed for free-space channels and several demonstrations have been performed (see \cite{Sidhu:Review2021,belenchia2021quantum} for review). The successful launch of the Micius QKD satellite in 2017 and the follow-up experiments \cite{Liao:Nat2017,Liao:PRL2018,Liao:NatPhoton2017,Ren:Nat2017}, have particularly been pivotal.}

{Free-space QKD systems must fight the effects of loss and noise in the link. For instance, a satellite-to-ground link would also encounter additional problems due to atmospheric turbulence and pointing errors. Such issues have been addressed widely through studying fading channels  \cite{Usenko:NJP2012,Papanastasiou:Fading2018}, analysing FSO QKD protocols \cite{Wang:AtmosNJP2018,Derkach:NJP2020,Derkach:Entropy2020,Dequal:SatCVQKD2021,Gyongyosi:2020}, and applying adaptive optics techniques, e.g., to suppress noise \cite{Chai:NJP2020,Zheng:SDandAOR2017,Cao:Adaptive2017}. 
In the same direction, by focusing on the establishment of quantum communication and QKD links, probability distribution functions (PDFs) of the transmittance for slant propagation paths were derived, and models for atmospheric quantum channels with turbulence were proposed \cite{Vasylyev:EllipticModel,Vasylyev:Elong2019}. 
In addition, distant FSO atmospheric channels have been experimentally characterized  \cite{Liao:Nat2017,Gunthner:GEOSat2017}, where optical loss and signal noise are measured. As well, attempts were made to stabilize transmittance fluctuations caused by beam wandering over free-space atmospheric channels \cite{Usenko:FSOptics2018}. }  

{On the other hand,} it is desirable to find the limits of quantum communications and QKD in different types of free-space medium, such as the Earth's atmosphere and space. In fact, alike the PLOB bound \cite{Pirandola:PLOB17} and quantum repeater capacities \cite{Pirandola:EndtoEnd2019}, one may work out bounds germane to free-space and satellite links, where the most detrimental phenomena is perhaps, not surprisingly, turbulence---fluctuations in the atmosphere refractive index due to the aerodynamics and temperature gradient of the Earth’s surface~\cite{Andrews:Book,Kaushal:Book}. 
Due to atmospheric turbulence the spatial coherence of an optical beam is gradually destroyed as it propagates. This loss of spatial coherence restricts the reach to which beams can be focused or collimated~\cite{Goodman:Book,Siegman:Book,Svelto:Book}. This in turn results in significant power level reductions in FSO communication and radar links. Equally fatal, the destruction of coherence can effect optical receivers, which are very sensitive to the loss of spatial coherence~\cite{Fried:1967,Murty:1979}.

Accounting for realistic effects on optical beams, such as diffraction, extinction, background noise, and channel fading, the latter due to pointing errors and atmospheric turbulence, Pirandola investigated the ultimate quantum communication limits and the practical security of FSO links, considering ground-based communications~\cite{Pirandola:FS2021} and uplink/downlink with satellites~\cite{Pirandola:Sat2021}. Even though the theory developed in \cite{Pirandola:FS2021} is very general, the main focus was the regime of weak turbulence, suitable for short-range high-rate FSO links on the ground. Similarly, the main focus of \cite{Pirandola:Sat2021} was quantum communications with satellites within 1 radiant from the zenith position, so to enforce the regime of weak fluctuations.

In this manuscript, we extend the investigation to the regime of moderate-to-strong turbulence~\cite{Fante:1975,Mironov:1977,Yura:1973}, where optical waves can harshly be deformed and eventually broken up into multiple patches~\cite{Klyatskin:1972,Murty:1979}, such that one would observe a random multiplicity of spots distributed on the receiving aperture~\cite{Kerr:1973,Raidt:1975}. 
{Of main tools in studying free-space links in the presence of atmospheric turbulence are PDFs, such as log-normal, extended Huygens-Fresnel, and the recently proposed elliptic-beam models~\cite{Vasylyev:EllipticModel,Andrews:Book}. Such functions are beneficial to the estimation of, e.g., transmissivity of FSO channels. However, they can be cumbersome to handle, even numerically, and therefore restrictive for a theoretical account of the system. As one key contribution to the body of the field, considering the purposes of quantum communications and QKD, we put a lower bound on the transmissivity of atmospheric links that alleviates security analysis of such systems. Not only the bound is manageable, but also it can be used at all turbulence regimes. }
Next, in the more challenging regime of moderate-to-strong turbulence, we provide information-theoretic bounds for the maximum rates that are achievable for key generation and entanglement distribution. We then study the composable finite-size key rates that can be achieved by protocols of CV-QKD, showing the feasibility of this approach in moderate-to-strong FSO links. 

The considered stronger regime of turbulence occurs in long-distance free-space connections on the ground but also in communications with satellites at large zenith angles (beyond 1 radiant). When a satellite is close to the horizon, the optical path within Earth's atmosphere becomes long and turbulence becomes a major problem. At these angles, another problem is refraction, which creates an elongation of the atmospheric section of the path (and therefore further loss and turbulence occur). Accounting for all these adverse aspects, we bound the optimal performances and provide achievable key rates.

\bigskip
{\bf Results and Discussion}

We first present some preliminary aspects and physics of FSO communications in turbulent media. We shall use these in the rest of the paper in order to understand and establish both ultimate limits and practical security of quantum communications in a moderate-to-strong turbulent space.

\smallskip
{\bf Figure of merit for the strength of turbulence.}
Assume an optical-beam signal of wavelength $\lambda$ that propagates through a turbulent path of length $z$. 
As widely accepted \cite{Fante:1975,Andrews:Book,Murty:1979}, we introduce the Rytov number to be the figure of merit for the strength of turbulence.
Physically, the Rytov number, or Rytov variance, is a measure of the strength of light scintillations---fluctuations in received irradiance, or in the phase and amplitude of the light, resulting from propagation through a turbulent space \cite{Andrews:BookScint,Andrews:Book}.  
The dimensionless Rytov number is defined for a plane wave as follows \cite{Rytov:1937}
\begin{align}
\label{Rytov}
\sigma_{\rm Ry}^2=1.23C_n^2k^{\frac{7}{6}}z^{\frac{11}{6}}, 
\end{align}
where $k=2\pi/\lambda$ is the wavenumber and $C_n^2$ is known as the index-of-refraction structure constant, measuring the magnitude of the fluctuations in the index of refraction (the Rytov number for a spherical wave is $0.4\sigma_{\rm Ry}^2$). 
Note that the scintillation of an optical signal does not increase unlimitedly as predicted by Rytov approximation \cite{Rytov:1937}, but saturates for strong turbulence and long propagation links \cite{Murty:1979}. It can nevertheless still specify turbulence regimes. 

Values of $\sigma_{\rm Ry}^2<1$ refer to weak turbulent media, while $\sigma_{\rm Ry}^2 > 1$ indicate strong turbulence~\cite{Fante:1975}. The regime of intermediate turbulent media hence is lying around $\sigma_{\rm Ry}^2 \sim 1$. 
Rytov number is very much similar to the dimensionless Reynolds number \cite{Sommerfeld:1908}, $Re$, in fluid mechanics, where for a fluid flowing through a packed bed of particles $Re<10$ corresponds to a laminar flow, whereas $Re>2000$ indicates a turbulent stream \cite{Rhodes:Book}.
According to the Rytov number, the specification of turbulence regimes involves not just the index-of-refraction structure constant $C_n^2$, but a combination of this parameter, the beam's wavelength and the propagation path length. 

The positive power dependence of the Rytov number on path length $z$ implies that the medium is indeed expected to be highly turbulent at longer distances \cite{Murty:1979}.  
It is hence helpful to introduce another quantity which is relevant to the propagation distance, which is~\cite{Yura:1971,Yura:1973,Fante:1975}
\begin{align}
z_i=(C_n^2k^2\ell_0^{5/3})^{-1}.    
\end{align}
Parameter $z_i$ represents the propagation length at which the transverse coherence radius of the optical wave is comparable to the turbulence inner scale $\ell_0$. The parameter $\ell_0$, which is on the order of 1~mm, is a measure of the smallest distances over which fluctuations in the index of refraction are correlated. We will shortly discuss that $z_i$ defines the minimum valid distance for some relevant quantities in studying stronger turbulence media; that is, some equations are sound only for $z>z_i$. Fortunately, apropos equations can be found in the literature for $z<z_i$, where we may expect a moderate or strong turbulence space.  
It is worth mentioning that, in the regime of weak turbulence, a similar quantity, known as the spatial coherence radius $\rho_0=(\dot\iota C_n^2k^2z)^{-\frac{3}{5}}$, is introduced, where $\dot \iota=0.55~(1.46)$ corresponds to plane (spherical) waves~\cite{Andrews:Book}.  

\smallskip
{\bf Pure diffraction and optical loss in free space.}
A natural light's phenomenon is diffraction, which perennially spreads the wave's size while it propagates through free space. It also constantly increases the radius of curvature of the propagating beam \cite{Siegman:Book,Svelto:Book}. 
In our study, we start with a Gaussian beam, with initial field spot size $w_0$, carrier wavelength $\lambda$, and radius of curvature $R_0$. 
At distance $z$ of propagation, where a receiver is supposedly placed, free-space diffraction increases the beam's spot size to 
\begin{align}
\label{radius:z}
    w_z^2=w_0^2\Big[\Big(1-\frac{z}{R_0}\Big)^2 + \Big(\frac{z}{z_R}\Big)^2\Big],
\end{align}
with $z_R=\pi w_0^2/\lambda$ being the beam's Rayleigh length. 
A receiver with infinite radius would collect all the light. However, practically speaking, only a fraction of the light can be collected by a receiver with a realistic finite aperture with radius $a_R$. This defines the pure diffraction-induced transmissivity 
\begin{align}
\label{diff:trans}
    \eta_{\rm dif}=1-e^{-\frac{2a_R^2}{w_z^2}}, 
\end{align}
yet, in reality, this would not be the total loss in a turbulent atmosphere as we shall see below. 

\begin{figure}[t]
    \centering
    \vspace{-0.2cm}
	\includegraphics[scale=0.6]{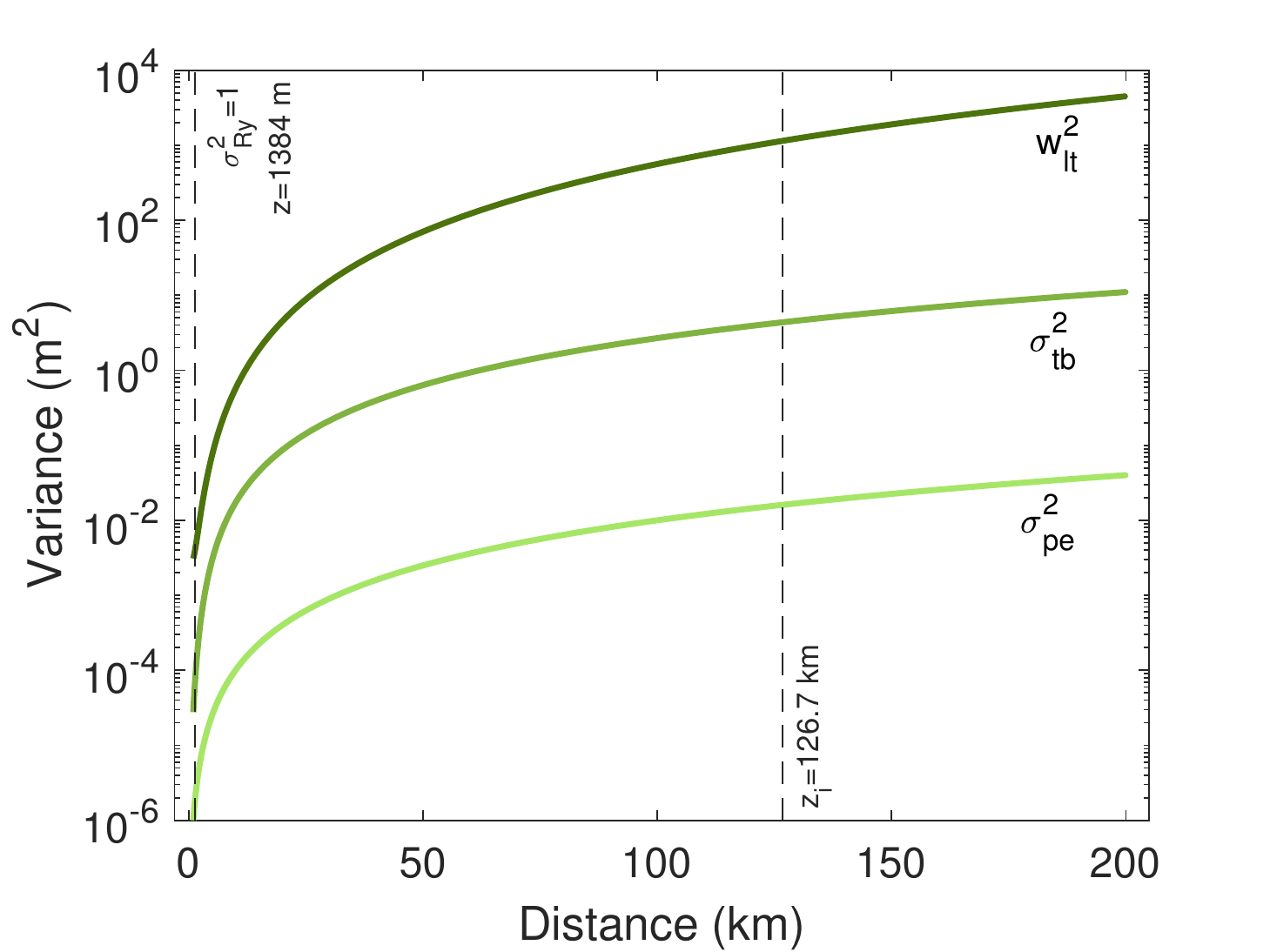}
	\caption{{\bf Beam widening in the presence of strong turbulence.} 
	Here, we compare the variance of the centroid wandering induced by turbulence ($\sigma_{\rm tb}^2$, middle line) to that of pointing error ($\sigma_{\rm pe}^2$, lower line) and the long-term beam waist ($w_{\rm lt}^2 $, upper line). 
	We assume a collimated beam ($R_0=+\infty$) with initial radius $w_0=5$~cm and wavelength $\lambda=800~\rm nm$. Other parameters are the outer scale of turbulence $L_0=1$~m and index-of-refraction structure constant $C_n^2=1.28\times 10^{-14}~{\rm m}^{-2/3}$ (night-time operation). Rytov variance ranges from $\sigma_{\rm Ry}^2 = 1$ at $z=1384$~m to $\sigma_{\rm Ry}^2 > 9.12\times 10^3$ at $z=200$~km.}
	\label{fig:PEvsBWvsBS}
\end{figure}

\smallskip
{\bf Turbulence-induced beam spread.}
Equation~\eqref{diff:trans} can lead to incorrect estimations because of Eq.~\eqref{radius:z}, which may underestimate the effective spot size of the beam.
This is because a different physics setting may apply in many real-world scenarios due to atmospheric turbulence.  
Therefore, we need to provide a proper estimation of the $z$-dependent spot size in order to modify $\eta_{\rm dif}$ in Eq.~\eqref{diff:trans}. 
In a moderate-to-strong turbulent regime, a beam can breakup into multiple patches and this primarily happens at longer propagation distances, where it is expected to have a large Rytov number. 
In this case, the patches of the beam will be in an area with mean square radius $w_{\rm lt}^2$, also known as the \emph{long-term beam waist} \cite{Murty:1979}. Note that the relevant beam spread in the regime of weak turbulence is the short-term beam waist, $w_{\rm st}^2$ \cite{Yura:1973}. In general, one has the decomposition $w_{\rm lt}^2=w_{\rm st}^2+\sigma_{\rm tb}^2$ \cite{Fante:1975,Yura:1973,Pirandola:FS2021}, where $\sigma_{\rm tb}^2$ is the variance associated with the wandering of the beam centroid. However, for stronger turbulence, wandering becomes negligible with respect to beam widening, i.e., we have the collapse $\sigma_{\rm tb}^2 \ll  w_{\rm st}^2\simeq w_{\rm lt}^2$. See Fig.~\ref{fig:PEvsBWvsBS} for a study of these quantities.

Let us now assume a Gaussian beam with initial spot radius $w_0$ and curvature $R_0$. After travelling through a path of length $z$, such a beam is characterized by a pair of parameters  \cite{Andrews:1994,Andrews:Book}
\begin{align}
\label{GBparameters1}
  \Omega_0= 1-\frac{z}{R_0}, ~  \Lambda_0=\frac{2z}{kw_0^2}. 
\end{align}
For example, the pair $\Omega_0=0$ and $\Lambda_0=0$ corresponds to a spherical wave, whereas $\Omega_0=1$ and $\Lambda_0=0$ represents a plane wave. Alternatively, in the plane of the receiver, such a Gaussian beam can be described by the similar pair of parameters
\begin{align}
\label{GBparameters2}
  \Omega=\frac{\Omega_0}{\Omega_0^2+\Lambda_0^2}=1+\frac{z}{R},~
  \Lambda = \frac{\Lambda_0}{\Omega_0^2+\Lambda_0^2}= \frac{2z}{kw_z^2},
\end{align}
where $R$ is the phase front radius of curvature at the receiver.  
It is then shown that, at distances $z>z_i$, where a strong turbulent space is experienced \cite{Murty:1979}, the long-term beam waist at the receiver is given by~\cite[Chap.~8]{Andrews:Book} 
\begin{align}
\label{radius:lt1}
w_{\rm lt} =w_z \sqrt{1+\frac{4}{3}q\Lambda},
\end{align}
with the $q$ parameter equal to 
\begin{equation}
q=0.74 \sigma_{\rm Ry}^2 Q_m^{1/6},~~Q_m=35.05 z/(k\ell_0^2).
\end{equation}
In Eq.~\eqref{radius:lt1}, we see how the diffraction-limited beam waist $w_z$ is revised into the long-term beam waist $w_{\rm lt}$ via an additional spread factor associated with scattering by turbulent eddies. 

Note that even through a short propagation distance the beam may experience a moderate or strong turbulence space. In this case ($z<z_i$) the effective beam waist is   
\begin{align}
\label{radius:lt2}
w_{\rm lt} =w_z \sqrt{1+1.63(\sigma_{\rm Ry}^2)^{\frac{6}{5}}\Lambda} .
\end{align}
The above equation is also considered to be adequately precise for weak turbulence so that it can generally be used to estimate the long-term beam waist under almost all turbulence conditions.
Thus, we may use Eq.~\eqref{radius:lt2} at all distances $0\simeq z<z_i$, no matter of the strength of turbulence. 

In this study, Eqs.~\eqref{radius:lt1} and \eqref{radius:lt2} provide the main quantities that we shall use to bound the rate of quantum communications in a moderate-to-strong turbulent space. 

\smallskip
{\bf More details on beam wandering.}
While transmitting an optical signal through free space, it is observed that position of the instantaneous centroid of the signal (point of maximum irradiance or “hot spot”) is randomly displaced. 
This instantaneous quivering in the plane of the receiver, which supposedly happens according to a Gaussian distribution with variance $\sigma^2$, is commonly called beam or centroid wandering.
Overall, this wandering is caused by pointing error $\sigma_{\rm pe}^2$, due to Gaussian jitter and off-target tracking, and atmospheric turbulence $\sigma_{\rm tb}^2$.
These two effects are independent and sum up such that the total variance of the wandering is given by $\sigma^2=\sigma_{\rm pe}^2+\sigma_{\rm tb}^2$. The amount of wandering for a typical $1~\mu$rad off-tracking error at the transmitter is given by $\sigma_{\rm pe}^2\simeq 10^{-12}z^2$. But, the contribution of atmospheric turbulence is more elaborate.

Different mathematical expressions have been developed to estimate wandering in strong turbulent media  \cite{Klyatskin:1972,Fante:1975,Andrews:Book,Mironov:1977}. 
Here, we use the following estimation  \cite[Chap.~8]{Andrews:Book}  
\begin{align}
\label{TBwandering}
\sigma_{\rm tb}^2 = 7.25 C_n^2 w_0^{-\frac{1}{3}} z^3  \int_0^1 d\xi \xi^2 \Bigg[\frac{1}{f^{\frac{1}{6}}(\xi)} - \frac{\kappa_0^{\frac{1}{3}} w_0^{\frac{1}{3}}}{\big[1+ \kappa_0^2w_0^2 f(\xi)\big]^{\frac{1}{6}} } \Bigg],
\end{align}
where $\kappa_0=2\pi/L_0$, with $L_0\simeq 1-100$~m being the outer scale of turbulence and
\begin{equation}
f(\xi)=[\Omega_0+(1-\Omega_0)\xi]^2 + 1.63 (\sigma_{\rm Ry}^2)^{6/5} \Lambda_0(1-\xi)^{16/5}.
\end{equation}
This is applicable in moderate-to-strong atmospheric turbulence, and is shown to be consisting with experimental data. 

As previously discussed, it turns out that centroid wandering is a negligible effect when turbulence is sufficiently strong. In Fig.~\ref{fig:PEvsBWvsBS}, we plot the turbulence-induced centroid wandering $\sigma_{\rm tb}^2$, the pointing-error wandering $\sigma_{\rm pe}^2$ and the long-term beam waist $w_{\rm lt}^2$.
While at short distances, where $\sigma_{\rm Ry}^2 \sim 1$, they tend towards each other, they diverge at longer distances, where $\sigma_{\rm Ry}^2 \gg  1$. 
Nevertheless, it is clear that at all distances considered, we have $w_{\rm lt}^2 \gg  \sigma_{\rm tb}^2 \gg \sigma_{\rm pe}^2$. In fact, the beam may break up into smaller patches in a very wide area, while the wandering of the centroid becomes negligible.

\smallskip
{\bf Turbulence-induced transmissivity.} 
In FSO communication, turbulence can cause power fading and sometimes complete loss of signal. 
In addition, communication links can experience severe signal degradation as well as spatial/temporal irradiance scintillations in the beam wavefront. 
To accurately estimate the signal fading and behaviour at some propagation distance, and to learn a true picture of how these affect crucial performance parameters such as the communication rate, it is important to analyze the distribution of the irradiance and/or transmittance at the receiver. 
In addition, having a theoretical distribution that accurately models these fluctuations under propagation conditions is desirable. 
This can be achieved through the knowledge of the statistical properties of the intensity fluctuations of the beams. In particular, the probability distribution of the transmittance most thoroughly characterizes the statistics of these fluctuations. Several models have been introduced to deal with this problem, including the log-normal model, the parabolic equation model, Feynman path integral, extended Huygens-Fresnel principle (see \cite{Andrews:Book}), and the recently proposed elliptic-beam model \cite{Vasylyev:EllipticModel}. 

The extended Huygens-Fresnel model is considered to be rather easier to use than other methods, especially when it comes to stronger turbulent media. 
For a Gaussian beam defined by the set of parameters given in Eqs.~\eqref{GBparameters1} and \eqref{GBparameters2}, and long-term waist given in Eqs.~\eqref{radius:lt1} and \eqref{radius:lt2}, the turbulence-induced transmissivity can be computed from 
\begin{align}
\label{etalt}
    \eta_{\rm lt}=  \frac{1}{\mathcal{N}}\int_{\mathcal{A}} d^2r \langle I(r, z) \rangle,  
\end{align}
where the integration is performed over the area $\mathcal{A}$ of the circular aperture, and 
\begin{align}
\label{etalt_N}
 \mathcal{N} = \lim_{\mathcal{A}\rightarrow \infty} \int_{\mathcal{A}} d^2r \langle I(r, z) \rangle
\end{align}
is a normalization factor. 
The mean irradiance $\langle I(r, z) \rangle$ is provided by the extended Huygens-Fresnel model \cite[Chapt.~7]{Andrews:Book}
\begin{align}
\label{meanIrrad1}
\langle I(r, z) \rangle = \frac{w_0^2}{w_{\rm lt}^2} \exp\bigg\{-\frac{2r^2}{w_{\rm lt}^2}\bigg\},   ~~~ z>z_i,
\end{align}
and 
\begin{align}
\label{meanIrrad2}
\langle I(r, z) \rangle = \frac{2w_0^2}{w_z^2} 
\int_0^\infty dt ~ t J_0\bigg(\frac{2\sqrt{2}r t}{w_z} \bigg) e^{-t^2-yt^{5/3}} 
,   ~~~ z<z_i,
\end{align}  
where $J_0(x)$ is a Bessel function and $y=1.41\sigma_{\rm Ry}^2 \Lambda^{\frac{5}{6}}$.

For $z>z_i$, we replace Eq.~\eqref{meanIrrad1} in Eqs.~\eqref{etalt} and \eqref{etalt_N}. Solving the integration, we can find an explicit analytical form for the transmissivity, given by 
\begin{align}
\label{eta_HF1}
      \eta_{\rm lt}=1- e^{-\frac{2a_R^2}{w_{\rm lt}^2}},
\end{align}
where $w_{\rm lt}^2$ is given in Eq.~\eqref{radius:lt1}.  
Thus Eq.~\eqref{eta_HF1} should be used instead of the pure diffraction transmissivity in Eq.~\eqref{diff:trans}. 

\begin{figure}[t]
    \centering
    \vspace{-0.2cm}
	\includegraphics[scale=0.6]{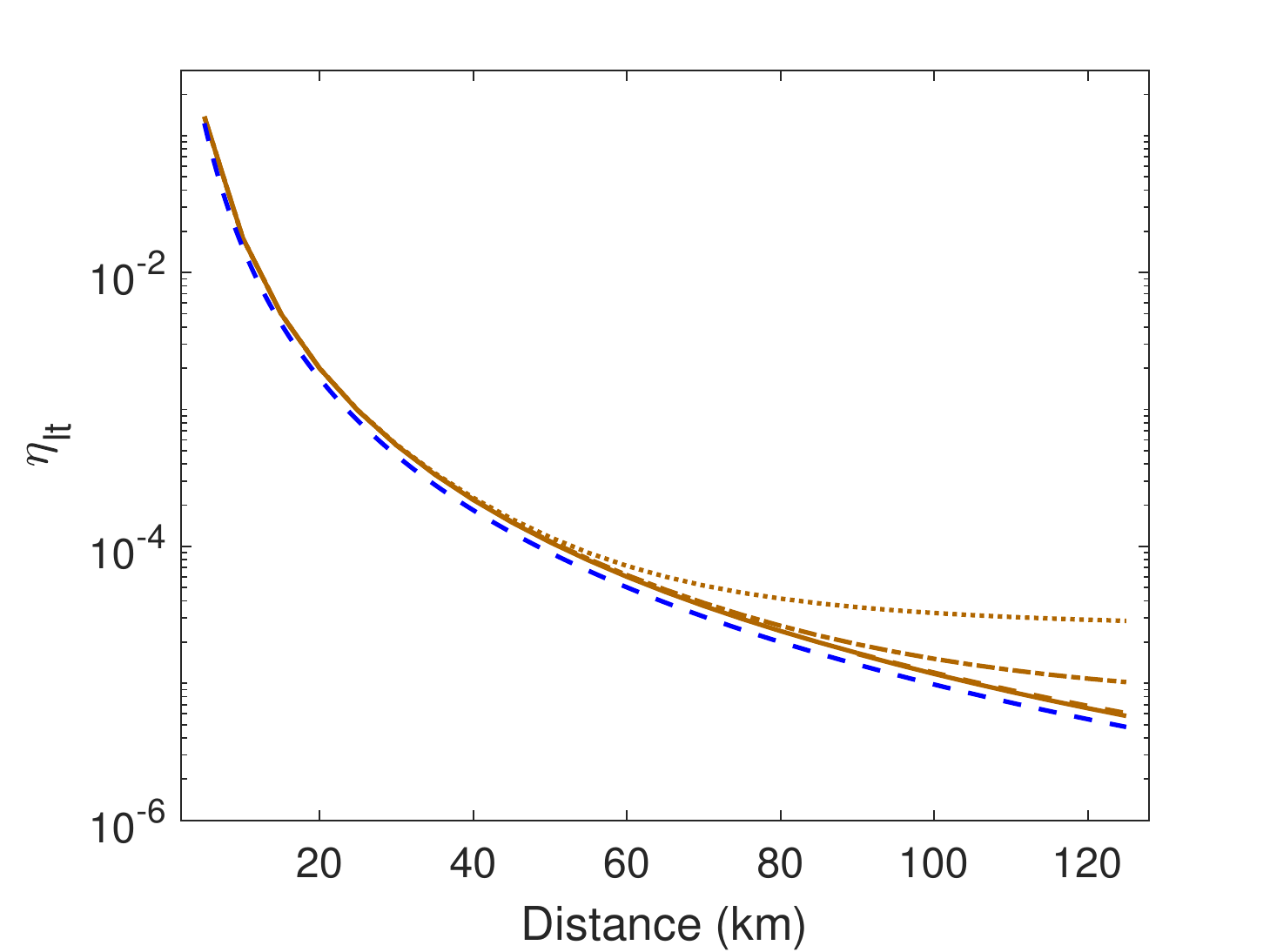}
    \caption{{\bf Turbulence-induced transmissivity versus distance.} By assuming the same parameters of Fig.~\ref{fig:PEvsBWvsBS}, here we plot turbulence-induced transmissivity versus distance $z<z_i$, where $z_i=126.7$~km. Brown curves from top to bottom correspond to the Huygens-Fresnel long-term transmissivity numerically computed for $a_R^\infty=$10, 20, 50, and 100 m. The lower (dashed blue) curve is the long-term transmissivity analytically computed from Eqs.~\eqref{eta_HF1} and.~\eqref{radius:lt2}. The latter can be assumed as limiting lower value at all distances.}
    \label{fig:transmiss}
\end{figure}

For $z<z_i$, we cannot find a closed-form but nevertheless we can compute the result numerically by replacing Eq.~\eqref{meanIrrad2} in Eqs.~\eqref{etalt} and \eqref{etalt_N}, and noting that the limit for unlimited area $\mathcal{A}$ can be treated by assuming $a_R = a_R^\infty$ for sufficiently large $a_R^\infty$. Notwithstanding, we can check that the formula in Eq.~\eqref{eta_HF1}, where we replace the long-term waist of Eq.~\eqref{radius:lt2}, provides a limiting lower bound to such numerical values, as shown in  Fig.~\ref{fig:transmiss}. Thus, we may use an analytical expression for the turbulence-induced transmissivity at all distances, as given by Eq.~\eqref{eta_HF1} where we replace either Eq.~\eqref{radius:lt1} (for $z>z_i$) or Eq.~\eqref{radius:lt2} (for $z<z_i$).

Another theoretical model is the log-normal model, where the beam follows a log-normal distribution rather than a Gaussian one. Using this model, we get a similar formula 
\begin{align}
\label{LNtrans}
    \eta_{\rm lt,LN} = 1- e^{-\frac{2a_R^2}{w_{\rm lt,LN}^2}},
\end{align}
where $w_{\rm lt,LN}^2$ is given in Methods. The validity of the formula holds for all propagation values $z$ and it has been experimentally verified~\cite{Capraro:LogNormalEXP}. 
In addition, it is shown to match recently developed descriptions of atmospheric transmissivity, such as the elliptic-beam model \cite{Vasylyev:EllipticModel}. However, the computation of $w_{\rm lt,LN}^2$ is cumbersome to handle even numerically. An heuristic choice is to combine Eq.~\eqref{LNtrans} with the calculation of the beam waist from other models, in particular, from the previous Huygens-Fresnel model. Thus, we may consider a {\em hybrid} log-normal model where we replace $w_{\rm lt,LN}^2$ with $w_{\rm lt}^2$, whose expression is given in Eqs.~\eqref{radius:lt1} and \eqref{radius:lt2}. This is completely equivalent to the previous approach. For this reason, in our study, we consider $\eta_{\rm lt}$ of Eq.~\eqref{eta_HF1} with long-term waist $w_{\rm lt}$ given by Eqs.~\eqref{radius:lt1} and \eqref{radius:lt2}.

\smallskip
{\bf Bounds and security of quantum communications in a moderate-to-strong turbulent space.}
Now we are in a position to account for the overall optical loss that can occur in a strong turbulence regime. The overall transmissivity includes the multiplication of three types of optical transmissivity
\begin{align}
\label{tot:loss}
\eta=  \eta_{\rm lt} \eta_{\rm eff} \eta_{\rm atm},
\end{align}
where we include the receiver's efficiency $\eta_{\rm eff}$  and atmospheric loss $\eta_{\rm atm}$.
The latter is modelled by the Beer-Lambert equation 
\begin{align}
  \eta_{\rm atm} =\exp\big\{-\alpha(\lambda,h_0) z\big\},~~\alpha(\lambda,h_0)=\alpha_0(\lambda) e^{-\frac{h_0}{6600}},
\end{align}
where $h_0$ is the altitude (measured in metres) and $\alpha_0(\lambda)$ is the extinction factor at sea level~\cite{Duntley:1948,Bohren:Book}.  

By replacing the combined transmissivity of Eq.~\eqref{tot:loss} in the repeaterless PLOB bound $\Phi(\eta)=-\log_2(1-\eta)$ \cite{Pirandola:PLOB17}, one gets the following upper bound for the rate $R$ of any QKD protocol over the FSO link 
\begin{align}
\label{UpBound}
R \leq \Phi(\eta) := -\log_2\Big[ 1-\eta_{\rm eff} e^{-\alpha(\lambda,h_0) z} \big(1- e^{-\frac{2a_R^2}{w_{\rm lt}^2}} \big) \Big].
\end{align}
We remark that, as shown in Fig.~\ref{fig:PEvsBWvsBS}, in the moderate-to-strong turbulence regime ($\sigma_{\rm Ry}^2 \geq 1$) the variance of long-term beam widening is several orders of magnitude larger than that associated with the centroid wandering. Therefore, we can neglect the short-term fading process and assume a fixed transmissivity between the sender and the detector plane at each distance.  
This is different from the weak turbulence regime where beam widening and wandering are equally important \cite{Pirandola:FS2021}. 

Apart form loss, the other key element that must be considered in FSO quantum communications is the number of thermal noise photons, which may find their way into the receiver's aperture.
They come from the sky brightness and can also be generated within the receiver itself. 
To involve the effect of thermal noise into the communications bound, we follow and apply the technique introduced in \cite{Pirandola:FS2021}.

The receiver sees a total mean number of thermal photons equal to $\bar n=\eta_{\rm eff}\bar n_B + \bar n_{\rm ex}$, where $\bar n_B$ and $\bar n_{\rm ex}$ are the number of background thermal photons per mode and extra photons generated within the receiver box, respectively.  
The number $\bar n_B$ depends on several factors coupled to the sky and the receiver. It is given by $\bar n_B=\pi \Gamma_R B_\lambda^{\rm sky}/\hbar\omega$, where $\hbar$ is the reduced Planck constant, $\omega$ is the angular frequency of light, and $B_\lambda^{\rm sky}$ is the brightness of the sky, which is in the range of $10^{-6}-10^{-1}~{\rm Wm^{-2}nm^{-1}sr^{-1}}$ from night to cloudy day \cite{Erlong:BackGNoise,Liorni:SatQKD2019}. 
The effects of the receiver is gathered in a single parameter $\Gamma_R=\Delta\lambda \Delta t \Omega_{\rm fov}a_R^2$, where $\Omega_{\rm fov}$, $\Delta\lambda$, and $\Delta t$ are the angular field of view, spectral filter, and time window of the detector, respectively. 
The nominal values that we use in this study are $\Omega_{\rm fov}=10^{-10}~{\rm sr}$, $\Delta\lambda=0.1~{\rm pm}$, and $\Delta t=10~{\rm ns}$. The natural interferometric effect of coherent detection, where the signal and LO pulse overlap, imposes an effective filter of $\Delta\lambda=\lambda^2\Delta\nu/c$, such that assuming $\lambda=800$~nm, a LO of $\Delta t=10$~ns, and a bandwidth $\Delta\nu=50\geq 0.44/\Delta t$~MHz, applies an effective filter of $\Delta\lambda=0.1$~pm.  
This would suppress the background noise $\bar n_B$ to the order of $10^{-12}$ ($10^{-7}$) at night (day) time, which in turn allow for positive rates that could not have been obtained otherwise. Precisely, for a receiver with $a_R=5~{\rm cm}$, we estimate $\bar n_B= 4.75 \times 10^{-12}~(10^{-7})$ background photons per optical mode at night (day). 

The total Alice-Bob FSO link is modelled as a thermal-loss channel with transmissivity $\eta$ and overall thermal noise $\bar n$. 
The worst-case scenario is when the eavesdropper (Eve) has control over all the input noise. Such a scenario can be simulated by her using a  beam splitter with transmissivity $\eta$ that combines Alice's signal mode with an input thermal mode with $\bar n_e=\bar n/(1-\eta)$ mean photons. 
We then use the thermal-loss version of the
PLOB bound. For $\bar n \leq \eta$, the secret key capacity in Eq.~\eqref{UpBound} can be revised to 
\begin{align}
\label{UpBound:TN}
   R \leq  K_{\rm UB}(\eta,\bar n) := \Phi(\eta) - \frac{\bar n}{1-\eta}\log_2\eta - h\Big( \frac{\bar n}{1-\eta} \Big),
\end{align}
where $h(x)=(1+x)\log_2(1+x)-x\log_2x$. 
One may also find the achievable lower bound given by the reverse coherent information \cite{GarciaPatron:RCI2009,Pirandola:RCI2009}, i.e., there is an optimal rate $R$ such that
\begin{align}
\label{AchLoBound}
R \geq K_{\rm LB}(\eta,\bar n) := \Phi(\eta) - h\Big( \frac{\bar n}{1-\eta} \Big).
\end{align}

\begin{figure*}
\includegraphics[width=\textwidth]{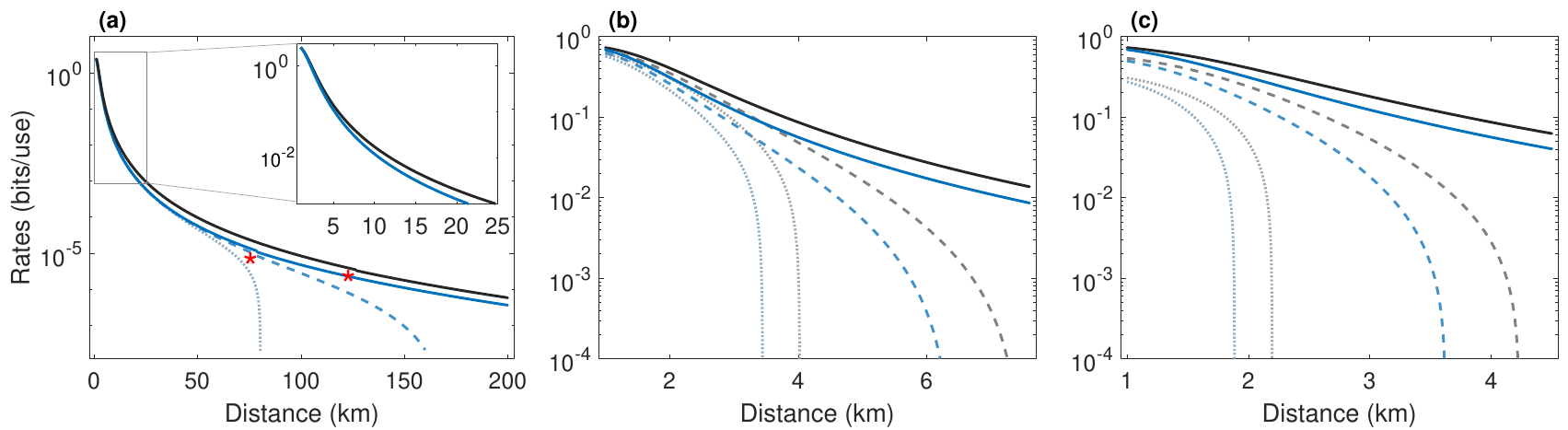}
\caption{{\bf Free-space optical quantum communications in a moderate-to-strong turbulent space.} We indicate night- and day-time conditions by black and blue curves, respectively. In (a) we plot the ultimate pure-loss bound of Eq.~\eqref{UpBound} with an ideal receiver, $\eta_{\rm eff}=1$ and $\bar n_{\rm ex}=0$, at night-time (solid black curves) and day-time (solid blue curve). The dashed (dotted) curves are thermal upper (achievable lower) bounds for an ideal receiver with $\eta_{\rm eff}=1$ but $\bar n_{\rm ex}=0$ [cf. Eqs.~\eqref{UpBound:TN} and \eqref{AchLoBound}]. 
The red star indicates the distance $z_i$ (connecting plots from different equations and therefore presenting small discontinuities). 
Here, the following set of parameters are considered: $\lambda=800~\rm nm$, $\alpha_0(\lambda)=5\times 10^{-6}~{\rm m}^{-1}$, $w_0=a_R=5~\rm cm$, $\Omega_{\rm fov}=10^{-10}~\rm sr$, $\Delta t=10~\rm ns$, $\Delta \lambda=0.1~\rm pm$, $h_0=30$~m, so that $C_n^2=1.28~(2.06)\times 10^{-14}~{\rm m}^{-2/3}$ for night (day). 
Also, we have thermal noise $\bar n_B=4.75\times 10^{-12}~(\times 10^{-7})$ photons per mode at night (day). 
In (b) and (c) we assume a lossy and noisy receiver with $\eta_{\rm eff}=0.5$ and, respectively, $\bar n_{\rm ex}=0.01$ and $\bar n_{\rm ex}=0.05$. 
As in panel (a), we compare the pure-loss rates (solid) with the thermal-noise bounds (dashed) and the achievable lower bounds (dotted). }
\label{fig:rate}
\end{figure*}

We present numerical simulations of the limits on communication rates in Fig.~\ref{fig:rate} showing the pure-loss bound of Eq.~\eqref{UpBound} and the thermal-loss bound of Eqs.~\eqref{UpBound:TN} and \eqref{AchLoBound}.
One first, and important, conclusion one may make is that we can obtain positive communication rates even in a strong turbulence regime. 

Each curve in Fig.~\ref{fig:rate}(a) is made of two parts because we have used two different equations in our simulation, i.e., Eq.~\eqref{radius:lt1} for $z \leq z_i$ and Eq.~\eqref{radius:lt2} for $z \geq z_i$. The distance $z=z_i$ is indicated by a red star, which is different for night and day operation (the right is for night). 
We observe a very slight inconsistency at $z=z_i$, which is due to using different expressions. Notwithstanding it is clear that the second part of the rate after $z_i$ follows exactly the same trend as the first part. 
In Fig.~\ref{fig:rate}(a) we compare the performances at night and day with an ideal receiver having $\eta_{\rm eff}=1$ and $\bar n_{\rm ex}=0$.  
For night-time operation all curves coincide because of absolutely low background noise ($\bar n_B=4.75\times 10^{-12}$). However, for day-time, with $\bar n_B=4.75\times 10^{-7}$, the deviation between the rates becomes distinct at large link distances, so that the thermal lower bound and upper bound drop at nearly 80~km and 150~km, respectively. Nevertheless the plot suggests that high rates can still be achieved at relatively shorter distances at both night and day.

Then we account for a realistic lossy and noisy receiver with $\eta_{\rm eff}=0.5$ and $\bar n_{\rm ex}=0.01$ in Fig.~\ref{fig:rate}(b), while  $\eta_{\rm eff}=0.5$ and $\bar n_{\rm ex}=0.05$ in Fig.~\ref{fig:rate}(c).
It is observed that the thermal photons generated at the receiver suppress the rates so that distances are of the order of a few kilometres. As we shall show later, this can be partially alleviated by using a receiver with a larger aperture size. 
 
Long free-space distances that we are considering here, e.g., $z=100$~km, may not seem so practical, especially because Earth's geometry, in particular its curvature, does not allow two terrestrial stations to actually ``see'' each other. For example, the maximum distance between two communications towers with height 30~m is about 40~km. Although this can be true for terrestrial stations, we allow for a wider variety of FSO links, including HAPS. 
Otherwise, a long-distance link could basically be an equivalent section of the atmosphere with a shorter length but stronger turbulence. 
 
The key rates for a moderate-to-strong turbulence regime can be seen as the tail of the rates  found in \cite{Pirandola:FS2021} for weak turbulence. This is where, at about 1384~m distance, we have $\sigma_{\rm Ry}^2=1$ and longer distances induce a stronger turbulence regime (for sake of comparison, we have used the same set of parameters used in \cite{Pirandola:FS2021}).
The main reason is that Eq.~\eqref{radius:lt2} is sufficiently precise even in weak turbulence regimes.
Let us also remark the reason behind choosing $\Delta \lambda=0.1$~pm, which is discussed in detail in \cite{Pirandola:FS2021}.

\smallskip
{\bf Composable finite-key security analysis.}
Equation~\eqref{AchLoBound} gives the achievable lower bound for key distribution rate when, ideally, an infinite number of signals are used for key extraction. 
However, in a real-world scenario, communication links can only be used a finite number of times. Hence, we may expect a poorer key rate than the asymptotic one. 
In addition, the security of a QKD protocol is desirable to be composable, i.e., the protocol must not be distinguished from an ideal protocol which is secure by construction \cite{Pirandola:AQCrypt}. Mathematically, a composable security proof can be provided by incorporating proper error parameters ($\varepsilon$'s) for each segment of the protocol, namely, error correction, smoothing, and hashing \cite{Tomamichel:2012,Furrer:FSize2012}.
To address this finiteness and composability, we study a QKD protocol based on coherent states for which we compute the composable finite-size key rate. 

We consider the homodyne-based coherent-state QKD protocol \cite{Grosshans:GG02PRL,Grosshans:GG02N}, the GG02 protocol, where Alice prepares $N$ Gaussian-modulated signals, with variance $V$, and sends them through a quantum channel to Bob. The latter performs a homodyne measurement, whereby he randomly measures one of the light quadratures. 
A number $n$ of signals will be used for key extraction, while the rest $m_{\rm pe}=N-n$ are left for parameter estimation. 
It can then be shown that the composable finite-size secret key rate is given by \cite{Pirandola:FS2021,Pirandola:Sat2021} 
\begin{align}
\label{comprate}
    R_\varepsilon  \geq & p_{\rm ec} (1-r_{\rm pe}) \Big( R_{\rm pe} - \frac{\Delta_{\rm aep}}{\sqrt{n}} + \frac{\Omega}{n} \Big),
\end{align}
where $p_{\rm ec}$ is the success probability of error correction connected to the frame error rate by $\textsc{FER}=1-p_{\rm ec}$, $r_{\rm pe}=m_{\rm pe}/N$ is the fraction of signals used for parameter estimation, 
$R_{\rm pe}$ is the asymptotic key rate accounting for parameter estimation, and \cite[Sec.~F]{Pirandola:CompCVQKD2021}
\begin{align}
    \Delta_{\rm aep}:= & 4\log_2(\sqrt{d}+2) \sqrt{\log_2(18 p_{\rm ec}^{-2} \varepsilon_{\rm s}^{-4} )} , \\
    \Omega:= & \log_2\big[p_{\rm ec}(1-\varepsilon_{\rm s}^2/3)\big] + 2\log_2(\sqrt{2}\varepsilon_{\rm h}). 
\end{align}

In Eq.~\eqref{comprate}, the asymptotic rate $R_{\rm pe}$ is calculated for the worst-case values of transmissivity and excess noise to be evaluated at the parameter estimation stage. 
These values are chosen within $w$ confidence intervals so that they are correct up to an error probability of $\varepsilon_{\rm pe}=\big[1-\text{erf}(w/\sqrt{2}) \big]/2$. See Methods for the calculation of $R_{\rm pe}$. 
 Equation~\eqref{comprate} is valid for a protocol with overall security $\varepsilon= \varepsilon_{\rm cor} +  \varepsilon_{\rm s} +  \varepsilon_{\rm h} + 2 p_{\rm ec}\varepsilon_{\rm pe}$ \cite{Pirandola:FS2021}, where $\varepsilon_{\rm h(s)}$ is the hashing (smoothing) parameter and $\varepsilon_{\rm cor}$ is the $\varepsilon_{\rm cor}$-correctness bounding the probability that Alice's and Bob's sequences are different even if they pass error correction. Finally, one needs to account 
for the analog-to-digital conversion so that each continuous-variable symbol is encoded in $d$ bits.

One further consideration regards the measurement techniques in CV-QKD. The received signals can be detected by using a coherent (homodyne or heterodyne) detection with the help of an either transmitted local oscillator (TLO) or local local oscillator (LLO). 
It turns out that at long distances the amount of detection noise is much lower for the LLO case. But, at the same time, the signal, which propagates through a turbulent path, and the LO, which is produced locally at the receiver, would not be spatially matched. As we show in Methods, this introduces even more loss to the system during the detection process. Therefore, we modify the overall transmissivity in Eq.~\eqref{tot:loss} by a further factor  $\eta_{\rm cd}$, i.e., 
\begin{align}
\eta=  \eta_{\rm lt} \eta_{\rm eff} \eta_{\rm cd} \eta_{\rm atm}.
\end{align}
Our estimate is that at long distances we roughly have $\eta_{\rm cd}=0.63$, which is the value used in our simulation. 

\begin{figure}
\includegraphics{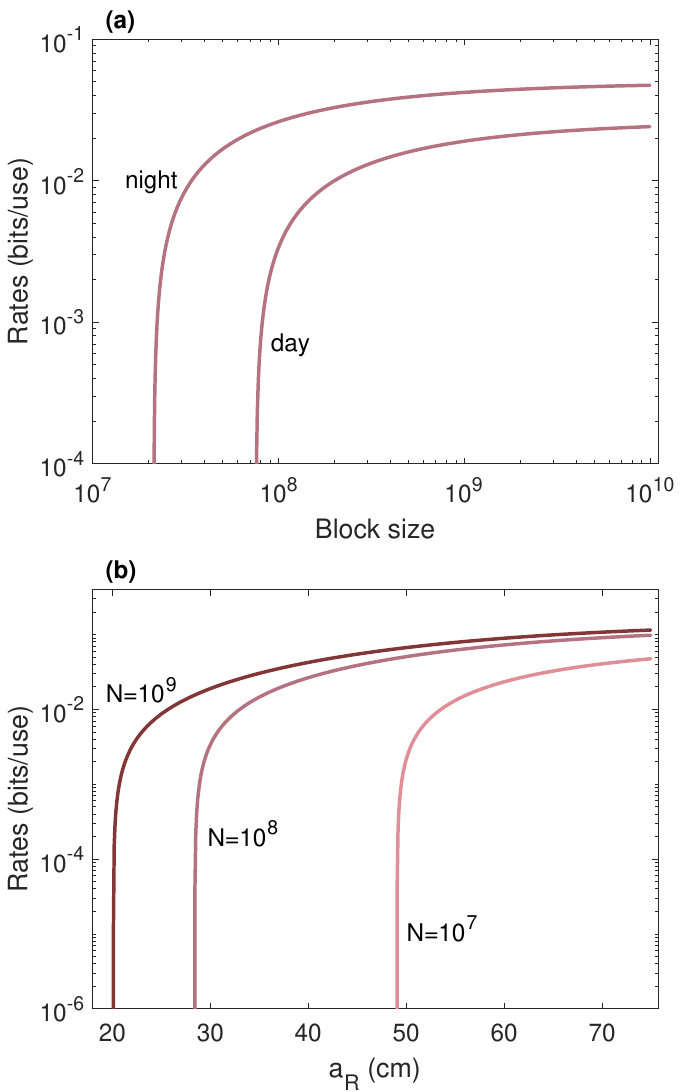}
\caption{{\bf Numerical results for the composable secret-key rate of a free-space CV-QKD protocol in turbulent space.} The rate is plotted versus (a) block size and (b) receiver aperture size. In (a), an aperture size $a_R=30$~cm is fixed.	In (b), assuming night-time operation, we plot the rate for various block-size values.  In both plots we assume a lossy and noisy receiver with $\eta_{\rm eff}=0.5$, $\eta_{\rm cd}=0.63$, and $\bar n_{\rm ex}=0.001$. Distance is $z=10$~km. Other physical parameters are set as given in Fig.~\ref{fig:rate}, except $\bar n_B$ which varies with $a_R$. Protocol parameters are: $\mu=10$, $r_{\rm pe}=0.1$, $d=2^5$, frame error rate ($\textsc{FER})$ is $0.1$, $\varepsilon_{\rm s}=\varepsilon_{\rm h}=\varepsilon_{\rm cor}=10^{-10}$, $w=6.34$, $\varepsilon=4.5\times 10^{-10}$, and $\beta=0.98$. }  
\label{fig:comprate}
\end{figure}

Fig.~\ref{fig:comprate} shows the composable finite-size key rate versus (a) block size and (b) receiver aperture size in a strong turbulence space. The link's length is $z=10$~km, equivalent to 7.84~dB, and the Rytov number is $\sigma_{\rm Ry}^2=37.56~(60.45)$ at night (day). 

In Fig.~\ref{fig:comprate}(a) we have fixed the receiver aperture size to $a_R=30$~cm. 
The rates at night-time operation can be obtained with a typical block-size of $\sim 10^8$, while the system demands a larger block-size, which is still acceptable. 
We observe that one main parameter that substantially affects the rates, at fixed distance and block-size, is the aperture size. 
From Fig.~\ref{fig:comprate}(b) we see that, at fixed length of $z=10$~km, positive rates can be achieved with a relatively large receiver. However, note that the aperture cannot be made too large.
In fact, increasing the receiver size lets more thermal photons into the detection system, e.g., we get $\bar n_B=1.71\times 10^{-10} (10^{-5})$ for $a_R=30$~cm, versus $\bar n_B=4.75\times 10^{-12} (10^{-7})$ for $a_R=5$~cm, at night (day).

\begin{figure*}
\includegraphics[width=\textwidth]{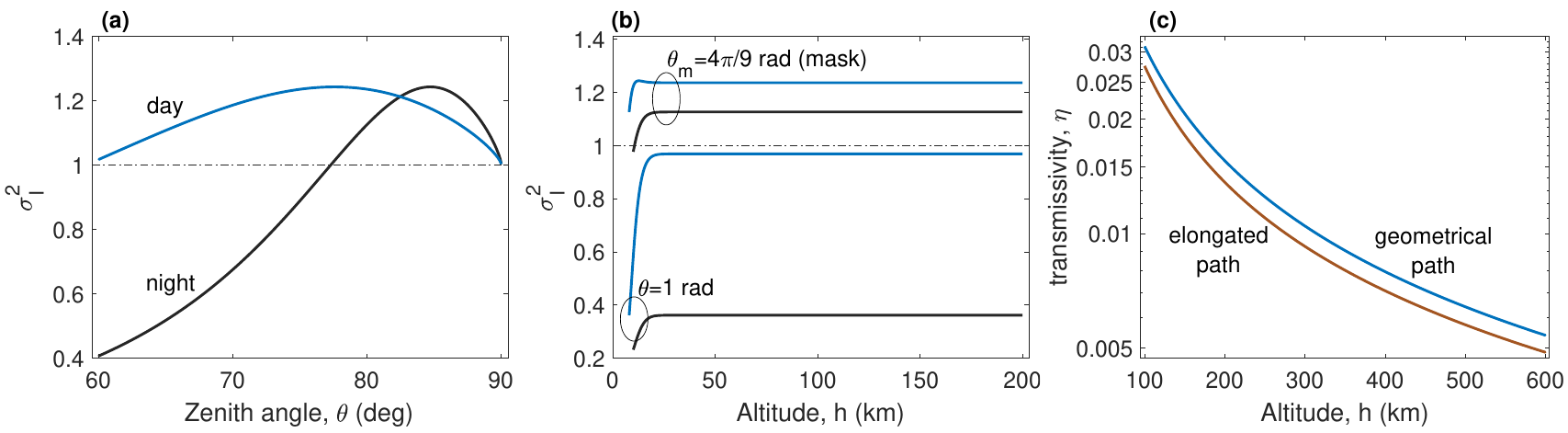}
    \caption{{\bf Satellite communications at large zenith angle.} In (a) we show the scintillation index of Eq.~\eqref{eq:ScintIndex} versus the zenith angle, at fixed $z=400$~km. In (b) we plot the scintillation index of Eq.~\eqref{eq:ScintIndex} versus altitude, at $\theta=1$~rad and $\theta_{\rm m}=4\pi/9$~rad. In (a) and (b) black curves are for clear-night turbulence conditions, while blue curves are for day-time and high-wind conditions. In (c) we illustrate the non-trivial difference between the elongated and geometrical paths at the mask angle $\theta_{\rm m}=4\pi/9$. In (c) we have set  $w_0=20$~cm, $a_R=40$~cm, $\lambda=800$~nm, $h_0=30$, $\alpha_0(\lambda)=5\times 10^{-6}~{\rm m}^{-1}$, and $\eta_{\rm eff}=0.5$.}
	\label{fig:Satellite}
\end{figure*}


\begin{figure}[t]
    \centering
    \includegraphics{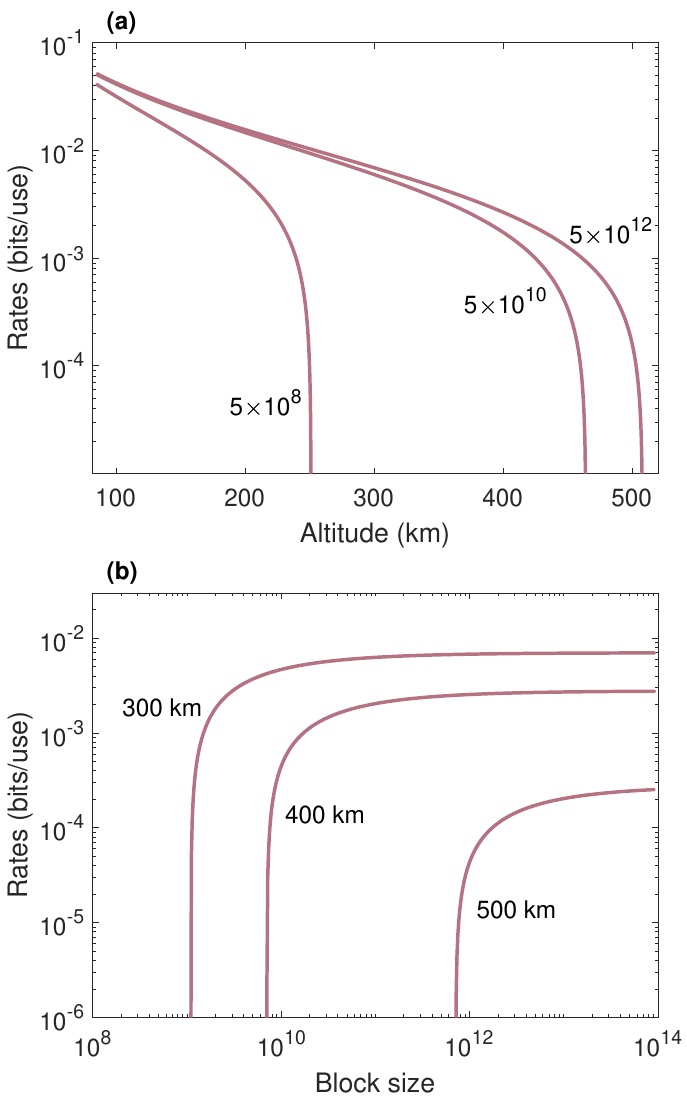}
    \caption{{\bf Performance of satellite quantum communications at large zenith angles.} In (a), we have finite-size key rates versus altitude (for fixed values of block-size). In (b) we have similar rates versus block-size (for fixed values of altitude). Both figures consider a mask angle $\theta_{\rm m}=4\pi/9$ at night-time, and windspeed $v=21~\rm m/s$ and $A=1.7\times 10^{-14}~\rm m^{-2/3}$ used in Eq.~\eqref{Cn2h}. Here we have set $w_0=20$~cm, $a_R=70$~cm, $\bar n_B=4.75\times 10^{-10}$, $\bar n_{\rm ex}=0.001$, and $\eta_{\rm cd}=0.63$. Other parameters chosen as given in Fig.~\ref{fig:Satellite}.  
    Protocol parameters are taken as follows: $\mu=10$, $\beta=0.98$, $r_{\rm pe}=0.1$, $d=2^5$, frame error rate ($\textsc{FER})$ is $0.1$, $\varepsilon_{\rm s}=\varepsilon_{\rm h}=\varepsilon_{\rm cor}=10^{-10}$, $w=6.34$, and $\varepsilon=4.5\times 10^{-10}$. }
	\label{fig:SatelliteRates}
\end{figure}

\smallskip
{\bf Satellite communications at large zenith angles.}
Here we apply the theory to a satellite communication link beyond 1~rad up to the horizon, where turbulence is strong. In particular, we focus on the mask (or cutoff) angle, $\theta_{\rm m}$, which is the minimum acceptable elevation above the horizon that a satellite has to be at to avoid blockage of line-of-sight. 
This is important because the key rates that will be derived for the mask angle represent lower bounds for the entire satellite quantum communication system.
One can set a mask angle that tells the receiver to ignore the satellite at zenith angles larger than $\theta_{\rm m}$, i.e., lower elevations.
The mask angle is roughly 80~deg ($4\pi/9$~rad) that is 10~deg from the horizon.
 
In this study, we consider a zenith-crossing satellite at altitude $h$, whose slant distance to the ground station, located at $h_0$ above sea-level, is given by 
\begin{align}
    z=\sqrt{(R_E+h)^2+(R_E+h_0)^2(\cos^2\theta-1)}-(R_E+h_0)\cos\theta,
\end{align}
where $R_E\simeq 6370$~km is Earth's radius and $\theta$ the zenith angle. 
To continue, we first need to identify the regime of operation. Replacing the above equation in the Rytov number of Eq.~\eqref{Rytov} cannot be used for a slant link out to the space because the index-of-refraction structure $C_n^2$ is not anymore constant and varies with the altitude $h$. We then require a more general, altitude-dependent, theory that stands as a measure for atmospheric scintillations and the turbulence regime. 
Assuming a downlink path from space, we take the following expression for scintillation index \cite{Andrews:2000} 
\begin{align}
   & \sigma_I^2(h,\theta)=  \nonumber \\ 
   & \exp \left[\frac{0.49\sigma_{\rm Ry}^2(h,\theta)}{\Big(1+1.11 \sigma_{\rm Ry}^{12/5}(h,\theta)\Big)^{7/6}} + \frac{0.51\sigma_{\rm Ry}^2(h,\theta)}{\Big(1+0.69 \sigma_{\rm Ry}^{12/5}(h,\theta)\Big)^{5/6}} \right] -1,
   \label{eq:ScintIndex}
\end{align}
where 
\begin{align}
  \sigma_{\rm Ry}^2(h,\theta)=2.25k^{\frac{7}{6}} \sec^{\frac{11}{6}}(\theta) \int_{h_0}^h dh'~ (h'-h_0)^{\frac{5}{6}} C_n^2(h').   \nonumber
\end{align}
In fact, $\sigma_I^2(h,\theta)$ is the modified version of a typical Rytov number that is now a function of altitude, zenith angle, as well as varying properties of the atmosphere.
According to the Hufnagel-Valley (H-V) atmospheric model \cite[Sec.~12.2]{Andrews:Book}, the index-of-refraction structure is a function of the altitude 
\begin{align}
\label{Cn2h}
    C_n^2(h)= & 5.94\times 10^{-53}(v/27)^2 h^{10} e^{-h/1000} \nonumber \\ & + 2.7\times 10^{-16}e^{-h/1500}+ Ae^{-h/100},
\end{align}
where $v$ is the windspeed~[m/s] and $A$ is the nominal value of $C_n^2(0)~\rm [m^{-2/3}]$ at the ground.  
In our simulation, we consider low-wind night-time by assuming $v=21~\rm m/s$ and $A=1.7\times 10^{-14}~\rm m^{-2/3}$, and high-wind day-time by assuming $v=57~\rm m/s$ and $A=2.75\times 10^{-14}~\rm m^{-2/3}$ \cite{Pirandola:Sat2021,Andrews:Book}. 

As it is seen in Fig.~\ref{fig:Satellite}(a), for zenith angles larger than 1 (1.32)~rad for day (night), we have $\sigma_I^2>1$, which means that signals will experience a moderate/strong turbulent space in such operational regimes. 
As $\theta\rightarrow 90$~deg scintillation drops to 1; precisely, to 1.0033. 
In addition, Fig.~\ref{fig:Satellite}(b) shows $\sigma_I^2$ versus altitude $h$, at the zenith angle $\theta=1$~rad as well as at the mask angle $\theta_{\rm m}=4\pi/9$~rad. 
At $\theta=1$~rad, the turbulence is weak for both night- and day-time operation, as also argued previously in \cite{Pirandola:Sat2021}. Whereas, at relatively high zenith angle, such as a mask angle of 80~deg, the turbulence in the link is strong at all values of altitude $h>20$~km. 

Another important factor that plays a role in a slant satellite path at large zenith angles is geometrical elongation of the communication
links. This is due to the refraction on interfaces of atmospheric layers, which introduces even more optical loss. 
It accounts for the apparent position of celestial objects toward the zenith, and is measured as the elongation factor, which is defined by the quotient of the (bent) optical trajectory and the (direct) geometrical slant path. 
We account for the elongation factor via the methodology introduced in \cite{Vasylyev:Elong2019}.
It uses the so-called standard atmosphere
model and distinguishes 10 atmospheric layers above the Earth's surface (within each layer the latitude dependence of refractive index is to be assumed linear). 
In Fig.~\ref{fig:Satellite}(c), we plot the optical loss for an elongated path, at night and at mask angle $\theta_{\rm m}= 4\pi/9$~rad, and compare it with that without elongation. It is seen that the elongated path imposes more optical loss.    

Let us now apply all the above consideration to the evaluation of finite-size key rates. 
In Fig.~\ref{fig:SatelliteRates}(a), for several block-size values, we have plotted key rates at night-time operation and at mask angle $\theta_{\rm m}=4\pi/9$~rad, 
where turbulence is strong (cf. Fig.~\ref{fig:Satellite}). 
Here we have set $w_0=20$~cm, $a_R=70$~cm, which constrains $\bar n_B=4.75\times 10^{-10}$, and $\bar n_{\rm ex}=0.001$.  
For the sake of comparison, we have also shown the pure-loss upper bound, which continue to offer higher rates with increasing the satellite altitude, whereas the finite-size rates drop at relatively lower altitudes. 
Furthermore, in Fig.~\ref{fig:SatelliteRates}(b), for several altitudes, we have plotted composable finite-size key rates versus block size, at night and at mask angle $\theta_{\rm m}=4\pi/9$~rad. 
Our simulation illustrates that with a reasonable block size and receiver size quantum satellite communication is feasible for altitudes up to 500~km.
At the same time, we note that the lifetime of low Earth orbit satellites with altitudes between 200 and 400~km is considerably short (fewer than three years) due to atmospheric drag, which eventually deorbits the satellites \cite{Cappelletti:Book}. This reads roughly 75 years for a satellite at 700~km altitude. 

{Finally, let us compare a part of our findings with actual measured data. For the Chinese Micius satellite \cite{Liao:Nat2017}, at altitude 500~km and zenith angle around 70~deg (that is a slant path of 1200~km), the loss was measured to be about 25~dB (using a transmitter telescope with 30~cm aperture size and a receiver telescope with 1~m aperture size placed at 890~m above ground level). There, with a repetition rate of 100~MHz, they could achieve a few kHz key rate from the satellite to ground by discrete-variable QKD protocols. This is comparable to our findings, at the same altitude and repetition rate, but a larger zenith angle (80~deg), which from Fig.~\ref{fig:SatelliteRates}(b) and at block size of $10^{12}$ reads 4.4~kHz key rate by CV-QKD protocols.
In addition, by assuming an Alphasat-like satellite in a LEO orbit at 500 km,  \cite{Gunthner:GEOSat2017} estimates the total channel losses from a satellite up to the receiving aperture, with an aperture of 1~m, to be about 20~dB (note that this is based on extrapolated data and not actual measured data). This is comparable to our results, read form Fig.~5c, that for the same orbit the channel loss is 16.4~dB. The difference may come from the choice of wavelength, which reads 1064~nm for their setup and 800~nm for ours, or the error in the extrapolation. }

In this work we have extended the field of FSO quantum communications
to a moderate-to-strong turbulent space where atmospheric conditions can be harsh and fatal to optical signals. Despite the possibility that the signals could be severely degraded and subjected to high optical loss, our results demonstrated that it is possible to obtain positive key rates.
After introducing a figure of merit for the strength of turbulence, we showed that in stronger turbulence regimes the beam spread dominates pointing errors and beam wandering, so that the latter effects can be ignored. 
We have then justified that the transmissivity estimated by a hybrid log-normal model can safely be used as a lower bound to the more elaborate extended Huygens-Fresnel model. 
 
With these tools in hand, we have computed the ultimate bounds for FSO quantum communication in moderate-to-strong turbulence regimes. 
Besides establishing these ultimate limits, we have also derived practical and composable finite-key rates for CV-QKD operated in such a strong turbulent space. An important feature is the level of excess noise generated at the receiver which may greatly reduce the key rates and reduce the distance for secure communication. However, our analysis also show that increasing the aperture of the receiver can mitigate the problem and revive the rates.  
As a main application of our results, we have then investigated satellite quantum communications at large zenith angles, specifically at the mask angle where not only turbulence is strong but also the 
elongation induced by refraction becomes relevant. This analysis allowed us to show that CV-QKD is feasible even in satellite links affected by strong turbulence, therefore removing the necessity and the restrictions
associated with the weak turbulence regime which is at the basis of previous literature.

\bigskip {\bf Methods}

We here present the main techniques that are needed to prove or support the results of our main text.

\smallskip
{\bf Transmissivity in a turbulence media: log-normal atmospheric model.}
In the log-normal model the probability distribution for the transmissivity is given by \cite{Vasylyev:EllipticModel} 
\begin{align}
    \mathcal{P}(\eta)= \frac{1}{\eta \sigma \sqrt{2\pi}} \exp\Big\{-\frac{(-\ln \eta - \mu)^2}{2\sigma^2}\Big\}, 
\end{align}
where $\mu=-\ln (\eta^2/ \sqrt{\langle \eta^2 \rangle})$ and $\sigma^2=\ln (\langle \eta^2 \rangle/ \eta^2)$ are parameters of the log-normal distribution. 
They are functions of the first and second moments of the transmissivity 
\begin{align}
\label{firstTM}
    \eta  =  \int_{\mathcal{A}} d^2 \bm r \langle I(\bm r, z) \rangle = \int_{\mathcal{A}} d^2 \bm r \Gamma_2(\bm r)
\end{align}
and 
\begin{align}
\label{secondTM}
    \langle \eta^2 \rangle & = \int_{\mathcal{A}} d^2 \bm r_1 d^2 \bm r_2 \Gamma_4(\bm r_1,\bm r_2),
\end{align}
where the integration is performed over the circular aperture opening area $\mathcal{A}$. In above equations, $\bm r=(x ~y)^T$ is the vector of transverse coordinates on the receiver plane. 

The field coherence functions $\Gamma_2$ and $\Gamma_4$ are respectively given by \cite{Vasylyev:EllipticModel} 
\begin{align}
    \Gamma_2(\bm r) = \frac{k^2}{4\pi^2 z^2} \int_{\mathbbm{R}^2} d^2 \bm r'
    e^{-\frac{g^2|\bm r'|^2}{2w_0^2} -2i\frac{\Upsilon}{w_0^2} \bm r.\bm r' - \frac{1}{2} D_S(0,\bm r')}
\end{align}
and 
\begin{align}
&  \Gamma_4(\bm r_1,\bm r_2) =  \frac{2k^4}{\pi^2 (2\pi)^3 z^4 w_0^2} \int_{\mathbbm{R}^6} d^2 \bm r_1' d^2 \bm r_2' d^2 \bm r_3' \nonumber \\
& \times e^{-\frac{1}{w_0^2}\big(|\bm r_1'|^2+|\bm r_2'|^2+|\bm r_3'|^2\big) } \nonumber \\
& \times e^{+2i\frac{\Upsilon}{w_0^2}  \big[ (1-z/R_0) \bm r_1'.\bm r_2' - (\bm r_1 - \bm r_2).\bm r_2'  - (\bm r_1+\bm r_2).\bm r_3'  \big] } \nonumber  \\ 
& \times \exp \Bigg[ \frac{1}{2} \sum_{j=1,2} \Big\{ D_S(\bm r_1-\bm r_2, \bm r_1'+(-1)^j\bm r_2')  \nonumber \\
& - D_S(\bm r_1 - \bm r_2, \bm r_1' +(-1)^j\bm r_3') - D_S(0,\bm r_2'+(-1)^j\bm r_3') \Big\}  \Bigg] ,
\end{align}
where $\Upsilon=kw_0^2/(2z)$ is the Fresnel number of the transmitter aperture and $g^2=1+\Upsilon^2(1-z/R_0)^2$ is the generalized diffraction beam parameter.
Here, 
\begin{align}
    D_S(\bm r, \bm r')=2\rho_0^{-5/3} \int_0^1 d\xi \big| \bm r \xi + \bm r' (1-\xi) \big|^{5/3} 
\end{align}
is the phase structure function, where $\rho_0$ is the radius of spatial coherence of the wave in the
atmosphere. 

The first moment of the transmissivity in Eq.~\eqref{firstTM} can be evaluated explicitly  
\begin{align}
    \eta = 1- e^{-\frac{2a_R^2}{w_{\rm lt,LN}^2}},
\end{align}
where  
\begin{align}
\label{LNlongterm}
    w_{\rm lt,LN}^2= & S_{xx}+4 \langle x_0^2  \rangle \nonumber \\
    \equiv & w_{\rm st,LN}^2 +\sigma_{\rm tb}^2
\end{align}
is the long-term beam size, with  
\begin{align}
    S_{xx}= & 4 \Bigg[ \int_{\mathbbm{R}^2} d^2 \bm r x^2 \Gamma_2(\bm r, z) \nonumber \\
    & - \int_{\mathbbm{R}^4} d^2 \bm r_1 d^2 \bm r_2  x_1 x_2 \Gamma_4(\bm r_1,\bm r_2, z) \Bigg]
\end{align}
and 
\begin{align}
    \langle x_0^2 \rangle = \int_{\mathbbm{R}^4} d^2 \bm r_1 d^2 \bm r_2  x_1 x_2 \Gamma_4(\bm r_1,\bm r_2, z).
\end{align}

\smallskip
{\bf Extra photons generated within the receiver.}
Considering a CV-QKD experiment, there are two techniques whereby one can measure the received signals through a coherent (homodyne or heterodyne) detection: transmitted local oscillator (TLO) and local local oscillator (LLO). In \cite{Pirandola:FS2021,Pirandola:Sat2021}, it is shown that these two may lead to generating totally different amounts of noisy photons within the coherent receiver system. This is mostly because extra photons generated by LLO, $\bar n_{\rm ex}^{\rm LLO}$, is a linear function of the link transmissivity, $\eta$, whereas extra photons generated by TLO, $\bar n_{\rm ex}^{\rm TLO}$, is an inverse function of it. 
Precisely, it reads \cite[Eq.~(62)]{Pirandola:FS2021} 
\begin{align}
    \bar n_{\rm ex}^{\rm LLO}=\Theta+\pi \eta V_A l_{\rm w} C^{-1} ~ \text{and} ~ \bar n_{\rm ex}^{\rm TLO}=\frac{\Theta}{\eta}, 
\end{align}
where  
\begin{align}
    \Theta=\frac{\nu_{\rm det}\textsc{NEP}^2 W\Delta t_{\rm LO} }{2\hbar\omega P_{\rm LO}},
\end{align}
with $V_A$ being the modulation variance, $P_{\rm LO}$ the LO power, $C$ the clock, $l_{\rm w}$ the linewidth, $W$ the detector bandwidth, \textsc{NEP} the noise equivalent power, $\Delta t_{\rm LO}$ the LO pulse duration, and $\nu_{\rm det}$ the detection noise variance---$\nu_{\rm det}=1 (2)$ for a homodyne (heterodyne) measurement. We refer to \cite{Pirandola:FS2021} for more detail. 

\begin{figure}[t]
\centering
\vspace{-0.2cm}
	\includegraphics[scale=0.6]{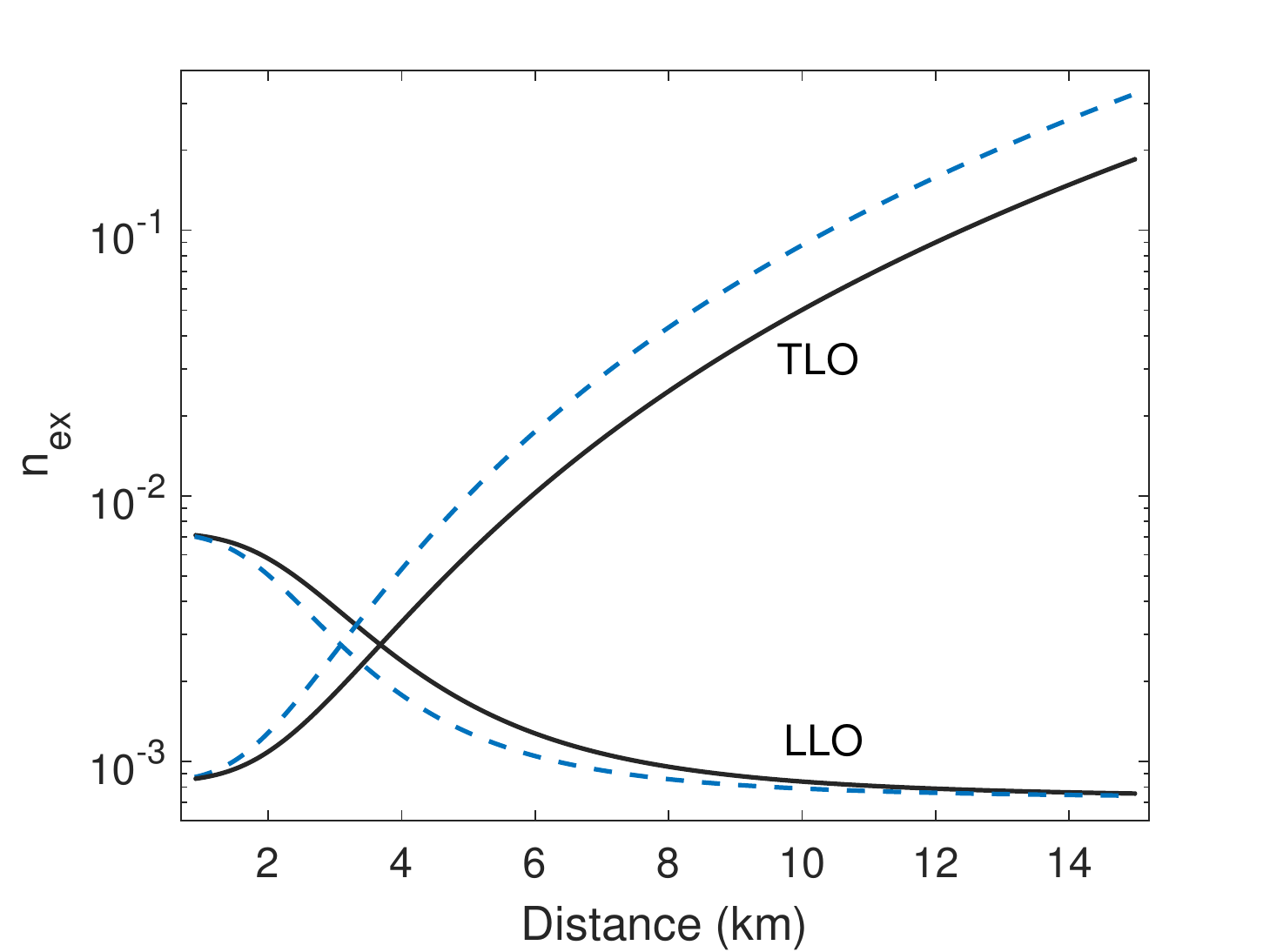}
    \caption{{\bf Extra noise photons generated within a coherent receiver (homodyne detection).} Here, we consider night- (solid black curves) and day-time (dashed blue curves) when a TLO/LLO technique is used. We have $\nu_{\rm det}=1$~SNU,
   $\textsc{NEP}=6~\rm pW/\sqrt{\text{Hz}}$, 
   $W=100$~MHz,
   $\Delta t_{\rm LO}$=10~ns,
   $P_{\rm LO}=100$~mW,
   $V_A= 8$~SNU,
   $l_{\rm w}=1.6$~KHz,
   $C=5$~MHz, and
   $hc=1.986\times 10^{-25}$~J.m. Other parameters related to $\eta$ are set as in Fig.~\ref{fig:rate}.}
    \label{fig:n_ex}
\end{figure}

In Fig.~\ref{fig:n_ex}, we plot $\bar n_{\rm ex}$ versus distance. As seen at relatively large distances, i.e., the regime of strong turbulence, the LLO technique is the better detection scheme.  
However, the quality of LLO detection may be poorer due to overlapping a fresh LO with the signal. In TLO, both the signal and the LO undergo the same (atmospheric turbulent) conditions, so that when they are recombined at the receiver, ideally, no mismatch is expected. 
This is not the case of LLO which we discuss in more detail in the following.

\smallskip
{\bf LLO-induced loss.} 
Suppose two continuous wave optical beams---the signal $E_S$ and the LO $E_L$---of the same frequency are incident on a beam splitter $\tau$. 
Let us consider a balanced homodyne detection, i.e., $\tau=1/\sqrt{2}$, where the output number of photons is given by \cite{Raymer:1995,Fried:1967} 
\begin{align}
\label{DiffPhotonNum}
    n_{-}= & \eta_{\rm eff} \int_0^T dt \int_{\mathcal{A}} d^2r [E_L^{-}(r,z,t) E_S^{+}(r,z,t) \nonumber \\
    & +  E_S^{-}(r,z,t) E_L^{+}(r,z,t)],
\end{align}
with spatial-temporal modes defined as follows 
\begin{align}
    E_S^{+}(r,z,t)=i\hat{a}_S f_S(t)u_S(r,z),  E_L^{+}(r,z,t)=i\hat{a}_L f_L(t)u_L(r,z),
\end{align}
and $\hat{a}$ being the corresponding annihilation operator.

Usually, for quantum tomography purposes and phase-sensitive detection, the LO field is assumed a monochromatic coherent state, with the on-axis amplitude $|\alpha_L|$, $f_L(t)=e^{-i\omega t}$, and $u_L(r,0) e^{i\phi_L}$ (plane wave) or $u_L(r,0)=e^{ikr}$ (spherical wave) \cite{Leonhardt:Book,Walls:Book,Fried:1967,Raymer:1995}. This then follows  
\begin{align}
    n_{-} \propto  \eta_{\rm eff} |\alpha_L|\big( \hat{a}_S e^{i\Delta \phi} + \hat{a}_S^\dagger e^{-i\Delta \phi} \big) = \eta_{\rm eff} |\alpha_L| \hat{q}_S(\Delta \phi),
\end{align}
where $\hat{q}_S(\Delta \phi)$ is signal's quadrature with $\Delta \phi=\phi_S-\phi_L$.

Back to the the coherent detection in a free-space scenario, in the following we show that some loss is expected in the case of LLO, where signal's shape is different from that of the LO.
We consider coherent Gaussian beams, which in the plane of the exit aperture of the transmitter are described by 
\begin{align}
u(r,0)= e^{-\frac{r^2}{w_0^2}-\frac{-ikr^2}{2R_0}}, 
\end{align}
where
$w_0$ is the beam spot radius and $R_0$ is its phase front radius of curvature. For simplicity, we assume a collimated beam with $R_0\rightarrow \infty$, such that  
\begin{align}
\label{GaussField}
u(r,0)= e^{-\frac{r^2}{w_0^2}}.
\end{align}
At distance $z$ a Gaussian beam may or may not keep its Gaussian form. If it does, the beam width $w_0$ will be replaced with $W(z)$---short- or long-term beam size according to the turbulence regime. 
However, in general, $u(r,z)$ can be distorted, or even completely destroyed, during a turbulent path. 
In that case, proper functions $u(r,z)$ should be used that reflect the effects of turbulence. 
We assume far-field conditions where Gaussian beams can be approximated by plane waves \cite{Andrews:Book}. Therefore, in the case of TLO, the signal and the LO can be taken as pane waves that reduces the problem to previous (usual) coherent detection scenarios \cite{Leonhardt:Book,Walls:Book,Fried:1967,Raymer:1995}, with the expectation value of photocurrent from Eq.~\eqref{DiffPhotonNum} as follows  
\begin{align}
    \langle n_{-}\rangle_{\rm TLO} \propto  \eta_{\rm eff} |\alpha_S(z)| |\alpha_L(z)| \cos(\Delta \phi). 
\end{align} 
When it comes to LLO, we should consider the Gaussian shape of the fresh LO generated locally at the receiver, while we assume the signal has the form of a plane wave. 
By replacing Eq.~\eqref{GaussField} for the LO into Eq.~\eqref{DiffPhotonNum}, and assuming that signal and the LO are frequency matched, it is straightforward to find   
\begin{align}
    \langle n_{-}\rangle_{\rm LLO} \propto \eta_{\rm eff} |\alpha_S(z)| |\alpha_L(0)| \cos(\Delta \phi) \frac{1}{\mathcal{N}_0}\int_{\mathcal{A}} dr~ r e^{-\frac{r^2}{W_L^2(0)}},
\end{align}
which is also normalized by $\mathcal{N}_0= \int_{\mathcal{A}\rightarrow \infty} dr~ r e^{-\frac{r^2}{W_L^2(0)}}$ (the receiver does not collect all the light). It is evident that the expression  
\begin{align}
    \eta_{\rm LLO}:= \frac{1}{\mathcal{N}_0}\int_{\mathcal{A}} dr~ r e^{-\frac{r^2}{W_L^2(0)}}
\end{align}
has the same nature as the quantum efficiency of the detectors $\eta_{\rm eff}$; hence, can be considered as extra loss. One can implicitly find that  
\begin{align}
    \eta_{\rm LLO}= 1- e^{-\frac{a_R^2}{W_L^2(0)}}. 
\end{align}
For the special case where the aperture size (or equivalently the lenses that collect and focus the beam on the detection's beam splitter) is equal to the LO's initial size, we have $\eta_{\rm LLO}=1-e^{-1}=0.63$.

The overall transmissivity can then be written as follows  
\begin{align}
\eta= \eta_{\rm lt} \eta_{\rm eff} \eta_{\rm cd}  \eta_{\rm atm},
\end{align}
where $\eta_{\rm cd}$ represents $\eta_{\rm TLO}$ or $\eta_{\rm LLO}$. In our estimation of composable CV-QKD rates, we use $\eta_{\rm cd}=0.63$. 

We remark that a more precise evaluation involves working out a more precise shape of the beam after propagating through a turbulent medium, where $u_{S/LO}(r,z)$ functions that include the effects of turbulent are known. 
One possible procedure is as follows: due to the extended
Huygens-Fresnel principle the optical wave field after propagating a distance $z$ through a turbulent space is given by solving \cite[Eq.~(21), Chapt.~7]{Andrews:Book}, where the most complex function seems to be the {\em complex phase perturbation} of the field \cite{Lutomirski:1971,Yura:1989}. 
One can then compute a more accurate loss coherent detection $\eta_{\rm cd}$ from theabove methodology.

\smallskip
{\bf Details of key rate analysis and parameter estimation.}
For the secret key rate analysis we use consider the entanglement-based representation of the coherent-state QKD protocol. We assume a collective Gaussian entangling-cloner attack \cite{Pirandola:CollecAttack2008}. 
At each run of the protocol Alice shares one leg of a two-mode squeezed vacuum (TMSV) state, with variance $\mu$, through a communications link with Bob. This is equivalent to the prepare and measure version of the protocol, where Alice prepares coherent states by a bivariate Gaussian modulation with variance $\sigma_x^2=\mu-1$. 
Assuming that the link is a thermal-loss channel, characterized by the transmissivity $\eta$ and thermal noise $\bar n$, the end-to-end covariance matrix between Alice and Bob has the form 
\begin{align}
\textbf{V}_{AB}=
\left(\begin{array}{cc}
a \mathbbm{1} & c \mathbbm{Z} \\
c \mathbbm{Z} & b \mathbbm{1}
\end{array}\right),
\end{align}
where $a=\mu$, $b=\eta(\mu-1)+2\bar n+ 1$, $c=\sqrt{\eta(\mu^2-1)}$, $\mathbbm{1}=\text{diag}(1,1)$ and $\mathbbm{Z}=\text{diag}(1,-1)$.

Having the triplet $(a,b,c)$, and assuming a homodyne measurement at Bob's side, the asymptotic key rate in the reverse reconciliation case is given by 
\begin{align}
\label{rate:asy}
R_{\rm asy}(\eta, \bar n)=\beta I_{AB}(\eta, \bar n)-\chi_{EB}(\eta, \bar n)
\end{align}
where  
\begin{align}
I_{AB}(\eta, \bar n)= \frac{1}{2} \log_2 \left(1+\frac{\eta(\mu-1)}{2\bar n+1}\right),
\end{align}
Also, assuming that the eavesdropper purifies the entangled state between Alice and Bob, one finds 
\begin{align}
\chi_{BE}(\eta, \bar n) = h\Big(\frac{\nu_+ -1}{2}\Big) + h\Big(\frac{\nu_- -1}{2}\Big) - h\Big(\frac{\nu_{\rm c} -1}{2}\Big). 
\end{align}
with $h(x)$ given in the main text, $\nu_\pm = \big(\sqrt{(a+b)^2-4 c^2} \pm (b-a)\big)/2$, and $\nu_{\rm c} =\sqrt{a( ab- c^2)/b}$.

In a realistic setting, Alice and Bob should compute the values of $\eta$ and $\bar n$ in order to estimate the key rate in Eq.~\eqref{rate:asy}. This computation is carried out by using only a finite number of runs, which inevitably reduces the rate to $R_{\rm pe}(\eta_{\rm wc}, \bar n_{\rm wc})$, for the worst-case values are $\eta_{\rm wc}\leq \eta$ and $\bar n_{\rm wc}\geq \bar n$ \cite{Ruppert:PE2014,Ruppert:PE2019}. 

Before discussing the worst-case scenario parameters, let us point out a matter that eases the parameter estimation in the case of moderate-to-strong turbulence. Unlike the case of a weak turbulence medium \cite{Pirandola:FS2021}, where the link transmissivity varies instantaneously, we can assume a fixed loss and a fixed number of thermal photons in the moderate-to-strong turbulence regime due to the fact that beam wandering is negligible here; see Fig.~\ref{fig:PEvsBWvsBS}. 
Therefore, we assume a thermal-loss channel that is characterised by transmissivity $\eta$ and mean number of thermal photons $\bar n$. This channel induces an input-output relation $y=\sqrt{\eta} x+z$ between the input Gaussian variable $x$ and the output variable $y$, with $z$ being a Gaussian noise variable; the variables $x$ and $z$ have zero mean with variances $\mu-1$ and $\sigma_z^2=2\bar n+1$, respectively. 

Back to the estimation of the worst-case parameters, by revealing $m$ pairs of corresponding data, i.e., $[x]_i$ and $[y]_i$, Alice and Bob can build an estimator $\widehat T$ of the square root of transmissivity $T=\sqrt{\eta}$, that is $\widehat T:=m^{-1}\sigma_x^{-2} \sum_{i=1}^{m} x_iy_i$, with variance $\text{Var}(\widehat T)=m^{-1}(2\eta + \sigma_x^{-2}\sigma_z^2)$, where $\sigma_x^2=\sum_{i=1}^{m} x_i^2\simeq \mu-1$. 
Then, the estimator for transmissivity is $\widehat \eta=(\widehat T)^2$, with variance $\text{Var}(\widehat \eta)=4m^{-1}\eta^2 \big(2+ \eta^{-1} \sigma_x^{-2} \sigma_z^2 \big)+\mathcal{O}(m^{-2})$. 
Similarly, Alice and Bob can construct the estimator for  $\bar n$, that is, $\widehat{\bar n}:=(\widehat{\sigma_z^2}-1)/2$, with variance $\text{Var}(\widehat{\bar n})=\sigma_z^4/(2m)$. Here, $\widehat{\sigma_z^2}=m^{-1}\sum_{i=1}^{m} z_i^2$ is the the estimator for the variance of the thermal noise $\sigma_z^2$. 

Next, by assuming a certain number $w$ of confidence of intervals, Alice and Bob compute the worst-case estimators up to some probability of error $\varepsilon_{\rm pe}=\big[1-\text{erf}(w/\sqrt{2}) \big]/2$, i.e., 
\begin{align}
    \eta_{\rm wc}= \eta - 2w\sqrt{\frac{2\eta^2 +\eta \sigma_x^{-2} \sigma_z^2 }{m}}, ~
    \bar n_{\rm wc}=  \bar n + w \frac{\sigma_z^2}{\sqrt{2m}}. 
\end{align}




\smallskip{\bf Acknowledgements.}
M.G. would like to thank Dmytro Vasylyev for helpful discussion regarding trajectory elongation. This work has been funded by the European Union via \textquotedblleft Continuous Variable Quantum Communications\textquotedblright\ (CiViQ, Grant Agreement No. 820466).


\bibliography{references}

\end{document}